\documentclass[pdflatex,sn-mathphys-num]{sn-jnl}

\usepackage[utf8]{inputenc} 
\usepackage[T1]{fontenc}    
\usepackage[semibold,tabular]{sourceserifpro}
\usepackage[charter,scaled=1.08]{newtxmath}
\usepackage[scaled=1.1]{zlmtt}
\usepackage[final,letterspace=25,nopatch=footnote]{microtype}
\usepackage{ragged2e}

\usepackage{amsmath,amsfonts}%

\usepackage[english]{babel}%

\usepackage{graphicx}%
\usepackage{multirow}%
\usepackage{mathrsfs}%
\usepackage[title]{appendix}%
\usepackage{xcolor}%
\usepackage{textcomp}%
\usepackage{manyfoot}%
\usepackage{booktabs}%
\usepackage{algorithm}%
\usepackage{algorithmicx}%
\usepackage{algpseudocode}%
\usepackage{listings}%
\usepackage[autopunct]{csquotes}%
\usepackage{setspace}%
\usepackage{xspace}%
\usepackage[
  mode = text,
  reset-text-family = false
]{siunitx}%

\newcommand{\CAPS}[2]{\textls[60]{\textsc{\MakeLowercase{#1}}}\textls[60]{#2}}
\newcommand{\caps}[1]{\CAPS{#1}{}}
\newcommand{\capsp}[1]{\CAPS{#1}{s}}
\newcommand{\DeePMD}{Dee\textls[70]{PMD}\xspace}
\newcommand{\NequIP}{Nequ\textls[70]{IP}\xspace}
\newcommand{\OpenMM}{Open\textls[70]{MM}\xspace}
\newcommand{\PySAGES}{\CAPS{P}{y}\caps{SAGES}\xspace}
\newcommand*{\emdash}{\;\rule[.5ex]{.5em}{.5pt}\;}
\newcommand{\osf}{\fontfamily{SourceSerifPro-TOsF}\selectfont}

\begin{document}

\title[Learning without Constraints]{
  The Importance of Learning without Constraints: Reevaluating Benchmarks for Invariant and Equivariant Features of Machine Learning Potentials in Generating Free Energy Landscapes
}


\author[1]{\fnm{Gustavo} \sur{Perez-Lemus}}\email{grp@uchicago.edu}
\author[1]{\fnm{Yinan} \sur{Xu}}\email{xuyinan@uchicago.edu}
\author[1]{\fnm{Yezhi} \sur{Jin}}\email{yzjin@uchicago.edu}

\author[1]{\fnm{Pablo} \sur{Zubieta Rico}}\email{pzubieta@uchicago.edu}

\author*[1]{\fnm{Juan} \sur{de Pablo}}\email{depablo@uchicago.edu}

\affil*[1]{\orgdiv{Pritzker School of Molecular Engineering}, \orgname{The University of Chicago}, \orgaddress{\street{5640 South Ellis Avenue}, \city{Chicago}, \postcode{60637}, \state{IL}, \country{USA}}}


\abstract{Machine-learned interatomic potentials (\capsp{MILP}) are rapidly gaining interest for molecular modeling, as they provide a balance between quantum-mechanical level descriptions of atomic interactions and reasonable computational efficiency. However, questions remain regarding the stability of simulations using these potentials, as well as the extent to which the learned potential energy function can be extrapolated safely. Past studies have reported challenges encountered when \capsp{MILP} are applied to classical benchmark systems. In this work, we show that some of these challenges are related to the characteristics of the training datasets, particularly the inclusion of rigid constraints. We demonstrate that long stability in simulations with \capsp{MILP} can be achieved by generating unconstrained datasets using unbiased classical simulations if the fast modes are correctly sampled. Additionally, we emphasize that in order to achieve precise energy predictions, it is important to resort to enhanced sampling techniques for dataset generation, and we demonstrate that safe extrapolation of \capsp{MILP} depends on judicious choices related to the system's underlying free energy landscape and the symmetry features embedded within the machine learning models.}

\keywords{machine-learning, enhanced sampling}



\maketitle

\setstretch{1.075}

\section{Introduction}\label{sec1}
In the field of molecular modeling, the underlying basis for accurate representation of a system is the availability of high-quality interatomic potentials or force fields. Development of such force fields often requires the use of quantum-chemical methods, such as density functional theory (\caps{DFT}). Due to their high computational costs, however, such approaches are generally limited to calculations on small systems and relatively few realizations of the system of interest.

Machine learning interatomic potentials (\capsp{MILP}) have emerged as a promising alternative for accelerated materials discovery and design. They offer a computationally efficient approach to achieve near \caps{DFT}-level accuracy at reduced computational costs. \capsp{MILP} are trained on datasets of atomic configurations that include energies and/or forces, from which the \caps{MILP} learns the potential energy surface (\caps{PES}) of the system. Some of the key differences between various advanced \caps{MILP} models reside in their approach to mapping the \caps{PES} into model-specific descriptors. Some models, such as \DeePMD\kern0em\cite{deepmd}, feature invariant atomic local descriptions, while others, like the Atomic Cluster Expansion (\caps{ACE}) used in systems such as Allegro\;\cite{allegro}, employ equivariant descriptors. \capsp{MILP} have been shown to be broadly applicable across diverse materials systems, ranging from atomic interactions in water\;\cite{signatures}, to silicon\;\cite{silicon}, to  catalysts\;\cite{nitrogen}.

From a practical perspective, \capsp{MILP} could be considered to be \textquote[]{black boxes}, thereby raising concerns about their reliability for prediction of physical or chemical properties\;\cite{short}. To address these concerns, several studies have sought to benchmark several \capsp{MILP} against a defined set of systems. These studies test and analyze directly the stability, physical correctness, and energy accuracy of a model through molecular dynamics simulations, rather than on the static datasets employed in traditional approaches\;\cite{validate,robust}.

Given the challenges associated with generating long molecular dynamics trajectories at the ab initio level, the evaluation of \caps{MILP} performance often relies on comparisons with results from classical forcefields. A notable example of a benchmark molecular system is the alanine dipeptide (\caps{ADP}) molecule. This molecule has a relatively simple structure, but offers a somewhat complex potential energy surface (\caps{PES}) that is governed by two principal torsional angles, phi ($\phi$) and psi ($\psi$)\;\cite{adp}.  For example, in\;\cite{forces}, the authors used \caps{ADP} in water as one of the benchmark systems. Similarly, in\;\cite{consider}, the authors used classical molecular dynamics to generate data sets for butane and \caps{ADP} in vacuum. In this latter study, identical atomic configurations were used to derive energy and force data at the \caps{DFT} level, thereby facilitating a comparison of \caps{MILP} behavior in these contexts. These two studies concluded that \capsp{MILP} face challenges in reproducing the free energy landscape of the $\phi$ and $\psi$ angles, and in their stability over prolonged molecular dynamics simulation times ($>$ 1 nanosecond of simulation).

The reported challenges of \capsp{MILP} in the \caps{ADP} benchmark necessitate a in-depth assessment of their to recreate the underlying physics embedded in the \enquote{source} potential across multiple benchmarks and over a range of conditions representative of typical applications. This assessment include careful consideration of the preparation of datasets (derived from classical molecular dynamics simulations) used to train the \capsp{MILP}. Such datasets must be representative of the wide range of ensemble configurations that a system might visit.

One crucial aspect to consider is the use of rigid-bond constraints, which are often employed in classical molecular dynamics simulations. These constraints are justified physically as they freeze the fast vibrational modes of the system without altering the overall dynamics, allowing for an increase in the timestep that can be used in classical \caps{MD} without compromising the numerical stability of a simulation\;\cite{constraints}. At a formal level, the use of algebraic constraints reduces the degrees of freedom of the simulated system by the number of constraints applied. Widely used algorithms to enforce constraints, such as \caps{SHAKE}\;\cite{shake}, \caps{SETTLE}\;\cite{settle}, or \caps{LINCS}\;\cite{lincs}, introduce constraint forces that must be considered in the development of a \caps{MILP}. 

This issue is relevant for the two \caps{ADP} benchmark examples mentioned above. In the work of\;\cite{forces}, the simulations used to generate the dataset for \caps{ADP} were performed with the \caps{LINCS} algorithm used on the hydrogen atoms in a solvated model of \caps{ADP}, whereas in the\;\cite{consider,consgithub} work, the \caps{SHAKE} algorithm was used to impose rigid constraints on the hydrogen-carbon bonds in vacuum. The training of \capsp{MILP} on these datasets raises several key considerations, which can be summarized by the following two points:
\begin{enumerate}
\item The forces in the datasets include a combination of conservative forces (the intrinsic gradient of the potential energy) and the constraint forces applied externally. Consequently, any model trained on both energy and forces encounter challenges in reconciling the differences between these forces, affecting learning performance. Even when models are trained solely on forces, as in the\;\cite{forces} study, the constraint forces cannot be represented through invariant or equivariant features. Consequently, the model lacks detailed information about the forces involved in the constrained bonds, thereby leading to numerical instabilities when the trained \caps{MILP} is used in \caps{MD} simulations. This can be observed in the simulations of the \caps{ADP} \caps{MILP} reported in\;\cite{forces}, where the instability begins with an incorrect distance between the hydrogen and carbon atoms of \caps{ADP}, the bonds that were constrained in the training dataset.


\item The datasets have the same relative distance for all the constrained atoms in every configuration of the ensemble, so even after recalculation of energy and forces with \caps{DFT}, as in\;\cite{consider}, the model sees the same local environment repeatedly, thereby failing to learn any bond interactions from corresponding to a given state. In this case, the bond energy of the \caps{MILP} model comes only from its inductive bias, and there is no guarantee that a stable and accurate \caps{PES} can be learned in that manner.
\end{enumerate}
In this work, we emphasize that the removal of mechanical constraints on target atoms when employing classical molecular dynamics (\caps{MD}) is essential for generation of training datasets for machine learning interatomic potentials (\capsp{MILP}). We select the \DeePMD and Allegro models and train them for \caps{ADP} in vacuum, in implicit solvent, and in explicit solvent (water). Our results show that model stability can be reached before \caps{PES} accuracy for both models with relatively small datasets. We find that the performance of the equivariant features of Allegro surpasses that of the invariant features in \DeePMD for a given dataset size. We also show that the use of enhanced sampling techniques allows for the accurate fitting of the \caps{PES} of the dihedrals of \caps{ADP} in vacuum and in an implicit solvent over the entire range of exploration. For the explicit solvent scenario, training solely with forces presents significant challenges for achieving model stability, though it does not correlate with \caps{PES} accuracy.

\section{Results}\label{sec2}
To evaluate the performance of \DeePMD and Allegro for the \caps{ADP} benchmark system, we rely on five distinct tests. We refer to the datasets obtained from unbiased simulations as ``direct'' dataset, whereas those obtained by enhanced sampling methods are referred to as ``enhanced'' datasets. The five tests are:
\begin{enumerate}
    \item Classical and \caps{DFT} \caps{PES} learning of \caps{ADP} in vacuum using direct simulations. \begin{equation}
        V_{\text{goal}}(r)=V_{\caps{ADP}}(r)
    \end{equation}
    where the $V_{\caps{ADP}}(r)$ is the same potential as that employed in\;\cite{consider}. The system is implemented in \caps{LAMMPS}\;\cite{lammps} with simulation monitoring performed using \PySAGES with the unbiased method\;\cite{pysages}. \PySAGES is a Python implementation of the Software Suite for Advanced General Ensemble Simulations (\caps{SSAGES})\;\cite{ssages} that provides full \caps{GPU} support for massively parallel applications of a variety of enhanced sampling methods.
    
    \item Classical and \caps{DFT} \caps{PES} learning of \caps{ADP} with enhanced sampling. \begin{equation}
        V_{\text{goal}}(r)=V_{\caps{ADP}}(r)
    \end{equation}
    In this test, we use \caps{ADP} mounted on \caps{SANDER}\;\cite{ambertools}, where the dynamics is modeled in \caps{ASE} coupled with \PySAGES. The enhanced sampling is achieved through the implementation of the Spectral~\caps{ABF}\;\cite{sobolev} method, which provides the best sampling scheme of the free energy surface (\caps{FES}) using functions in Sobolev spaces.
    
    \item Classical \caps{PES} Learning of \caps{ADP} with Implicit Solvent\;\cite{gbsa} using direct/enhanced sampling.
    \begin{equation}
        V_{\text{goal}}(r)=V_{\caps{ADP}}(r)+V_{\caps{GBSA}}(r)
    \end{equation}
    We use the same system as the one above, but include an implicit solvent potential ($V_{\caps{GBSA}}(r)$) at the \texttt{igb\,=\,1} level using \caps{SANDER}\;\cite{ambertools,gbsa}.
    
    \item Classical \caps{PES} learning of \caps{ADP} with explicit solvent using forces only
    \begin{equation}
        \nabla V_{\text{goal}}(r)=\nabla V_{\caps{ADP}}(r)+\langle \nabla V_{\text{eff}}(r)\rangle
    \end{equation}
    Here, $\langle \nabla V_{\text{eff}}(r) \rangle$ represents the average effective force exerted by the solvent on the \caps{ADP} molecule. This is the same test as the one detailed in\;\cite{forces}. We use \OpenMM\;\cite{openmm} coupled with \PySAGES and Spectral~\caps{ABF} for enhanced sampling.
    
    \item Classical \caps{PES} learning of the \caps{ADP} and water simultaneously
    \begin{equation}
        V_{\text{goal}}(r)=V_{\caps{ADP}\,+\,\text{water}}(r)
    \end{equation}
    For this last test, \OpenMM is used with \PySAGES for direct and enhanced sampling datasets generation; the enhanced dataset was generated with Spectral~\caps{ABF}.
\end{enumerate}

In all cases, constraints on hydrogen bonds in the \caps{ADP} (and water for test 5) molecule are removed, 1000 snapshots are collected for each simulation, and the datasets are randomly split into \qty{90}{\percent} for training and \qty{10}{\percent} for validation. In the first three sets, the direct calculations are controlled at \qty{500}{K}, whereas the enhanced distributions are calculated at \qty{300}{K}. For the fourth case, the enhanced simulations are carried out at \qty{300}{K} and the direct ones at \qty{600}{K}. For this same case, additional direct datasets with 40,000 snapshots were generated as well as enhanced datasets generated at \qty{300}{K}. For the last case, only two datasets were generated, 10,000 snapshots for direct simulation at \qty{600}{K} and 10,000 snapshots for enhanced sampling at the same temperature. For validation against the free energy surface and stability, \qty{5}{ns} enhanced sampling simulations were carried out with the \caps{MILP} implemented using \caps{ASE} coupling with \PySAGES (Spectral~\caps{ABF} method). Additional details of the simulations can be found in the Methods section. 

\subsection{Test 1}\label{subsec1}
\begin{figure}[htb]
\centering
\includegraphics[width=0.9\textwidth]{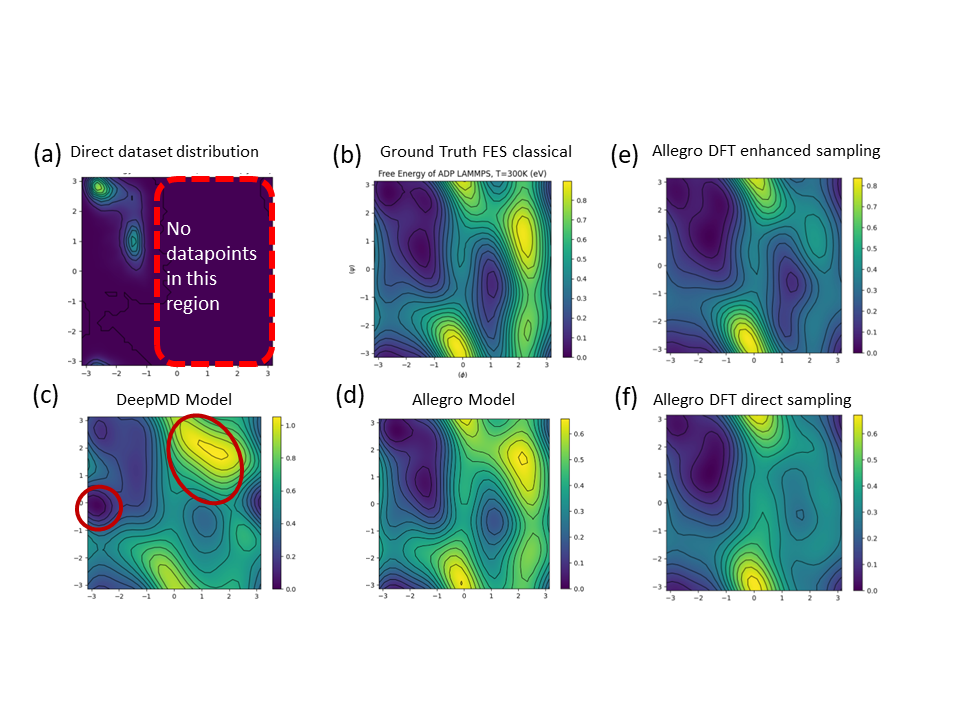}

\caption{Summary of Tests 1 results: a) histogram of the training dataset for Test 1 reflected on the dihedral angle space (the $\phi$ and $\psi$ angles), where no data points lie outside the two main energy minima of the system; b) the ground truth free energy profile of \caps{ADP} in vacuum using classical simulations; c) free energy profile using a \DeePMD model trained on the dataset shown in a), regions with notable discrepancies compared to the baseline are marked in red; d) free energy profile using an Allegro model trained on the dataset shown in a), where key energetic minima and the energy barriers are predicted well by the model. Summary of Test 2 results: e) free energy profile of \caps{ADP} at \caps{DFT} level with enhanced sampling; f) free energy profile based on the Allegro model trained on the dataset shown in (a) (the energies and forces are re-calculated using \caps{DFT}). The Spectral~\caps{ABF} method, executed in \PySAGES, is used for the free energy calculations in both tests.}\label{fig1}
\end{figure}

The findings from this test are illustrated in Figure \ref{fig1}. Figure \ref{fig1}a shows a histogram of the dataset projected onto the 2D space defined by the $\phi$ and $\psi$ angles, where the direct simulation predominantly explores the principal basin in the \caps{ADP} free energy surface (\caps{FES}). Fig. \ref{fig1}b shows the ground truth \caps{FES} of the model, which serves as the goal for Test 1. As demonstrated in Figures \ref{fig1}c and \ref{fig1}d, the \caps{MILP} models, trained using 1000 snapshots, are sufficiently robust to enable \qty{5}{ns} of enhanced sampling simulations. However, discrepancies are observed in some regions when using the \DeePMD model (Figures \ref{fig1}c) compared to the ground truth. Adding training data pertining to these regions could potentially improve the performance of the \DeePMD model and more accurately capture the \caps{ADP} \caps{FES}. In contrast, the Allegro model provides a high degree of accuracy across the entire \caps{FES}, with only minor deviations in the relative locations of the minima and maxima.

\subsection{Test 2}\label{subsec2}
The key results for Test 2 are shown in Figures \ref{fig1} (e--f), where, with a dataset calculated at the \caps{DFT} level, the Allegro model exhibits robust stability. The model using the enhanced sampling dataset yields an accurate prediction of the \caps{FES} and its key energy features. However, when trained on the direct sampling dataset, the Allegro model underestimates the unexplored energy minima compared to the performance with the classical dataset, as shown in Figure \ref{fig1}f. This discrepancy can be attributed to the inherently more complex potential energy surfaces (\capsp{PES}) at the \caps{DFT} level, which require a larger dataset for accurate predictions. Additional results in fig. \ref{fig2} (a and b) show the accuracy in the \caps{FES} of Allegro with enhanced sampling with classical data, and in fig. \ref{fig2} (c and d) how well Allegro predicts the energy directly on enhanced sampling simulations.

\subsection{Test 3}\label{subsec3}
\begin{figure}[ht]
\centering
\includegraphics[width=0.9\textwidth]{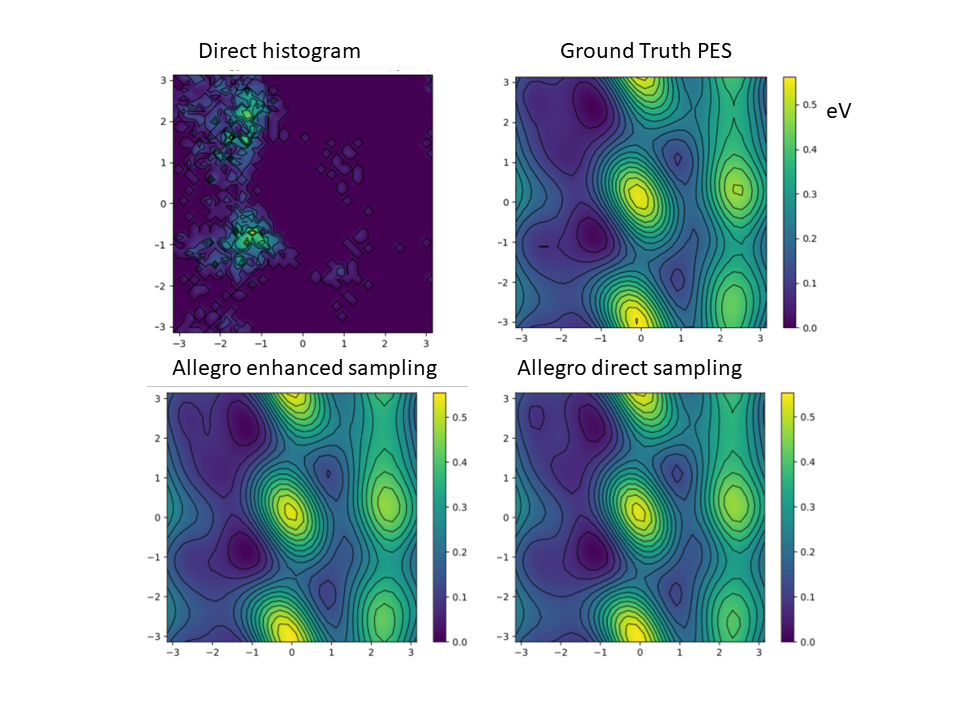}
\caption{Summary of Tests 3 results: (Upper left) histogram of the training dataset for Test 3 in the dihedral angle space (the $\phi$ and $\psi$ angles); (Upper right) the ground truth \caps{FES} of the \caps{ADP} system based on \caps{SANDER}; the \caps{FES} generated using Allegro model trained with enhanced (bottom left) and direct (bottom right) datasets. All \caps{FES} are generated using the Spectral~\caps{ABF} scheme.}\label{fig3}
\end{figure}

In implicit solvent simulations, Allegro generates stable potentials using a dataset of only 1000 snapshots, regardless of whether we use enhanced or direct data. As illustrated in Figure \ref{fig3}, Allegro achieves remarkably accurate results, independent of the nature of the training set. Here, in contrast to the results shown in fig. \ref{fig1}, a reasonable sampling of energy minima is sufficient to accurately capture the relative energies between them. Additionally, Allegro can predict the locations of maxima, even in the absence of direct data points at these maxima.

\subsection{Test 4}\label{subsec4}
\begin{figure}[htb]
\centering
\includegraphics[width=0.9\textwidth]{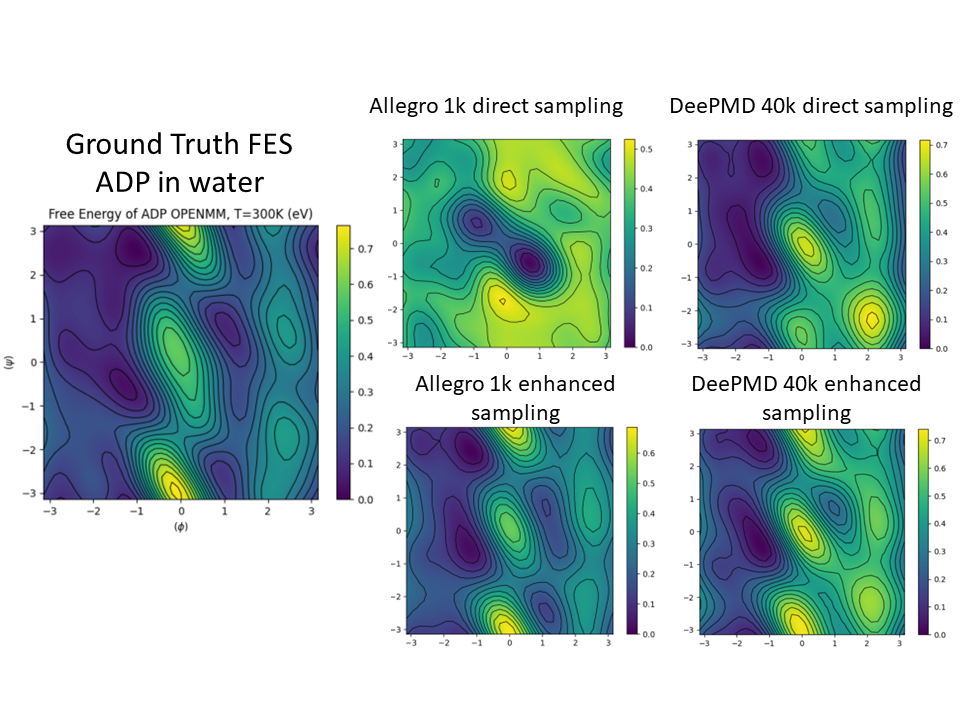}
\caption{Summary of Tests 4 results: (left) \caps{FES} of \caps{ADP} in TIP3P water in classical \caps{MD}; (top middle) the \caps{FES} of \caps{ADP} using the Allegro model trained on 1,000 snapshots from direct sampling at \qty{600}{K}; (bottom middle) \caps{FES} of \caps{ADP} using Allegro model trained on 1,000 snapshots taken from enhanced sampling at \qty{300}{K}; (top right) \caps{FES} generated with \DeePMD using 40,000 snapshots with enhanced sampling; (bottom right) \caps{FES} generated with \DeePMD using 40,000 snapshots with direct sampling. All \caps{FES} are generated using the Spectral~\caps{ABF} scheme.}\label{fig4}
\end{figure}

The results for this test are presented in Figure \ref{fig4}. The figures compare \caps{FES} Allegro models using direct and enhanced sampling and 1,000 snapshots for training to the \caps{FES} of \caps{ADP} with explicit water. As can be seen in the figure, the \caps{FES} from the model using a direct dataset does not agree with the ground truth, although it is stable throughout the 5 ns of simulation time with enhanced sampling (Spectral~\caps{ABF}). On the other hand, the Allegro model trained with enhanced sampling is able to reproduce the ground truth \caps{FES} with reasonable accuracy. When the dataset for both direct and enhanced sampling methods is increased to 40,000 configurations, the \DeePMD models exhibit higher stability, leading to accurate free energy profiles that are in good agreement with the ground truth, as shown in Figures \ref{fig4} (e and f). For the Allegro models, this increase in dataset size to 40,000 snapshots significantly refines accuracy (Fig. \ref{fig7}). The \caps{RMSE} for all tests are reported in Table \ref{table:rmse}.

To examine the impact of temperature on dataset preparation, we introduced an additional direct dataset containing 40,000 snapshots generated at \qty{300}{K}. This dataset is similar to that used in\;\cite{forces}, except that the hydrogen-bond constraints in \caps{ADP} were removed in our system. The results shown in fig. \ref{fig5} suggest that the model is unstable and fails to complete enhanced sampling calculations over a 5 ns trajectory. To investigate the cause of such instability, the \caps{PES} for the \caps{ADP} principal dihedrals is also calculated. This is achieved by passing a uniform distribution of \caps{ADP} conformations to the unstable model, calculating the potential energy, and averaging over a grid that covers the entire range of both dihedral angles. This approach aims to identify any deficiencies in the \caps{PES} that may account for the observed instability. The results, shown in Figure \ref{fig5}b, suggest that there is no gap in the \caps{PES} and, in fact, the model can predict the \caps{PES} with reasonable accuracy, a feature that the relatively stable \NequIP and GemNet models of\;\cite{forces} lack. Further analysis of the instability shows that the failure of the model is caused by the high proximity of two hydrogen atoms bonded to the nitrogen in the \caps{ADP}, resulting in excessively high forces that destabilize the system. This is shown in Figure \ref{fig5}c, where the potential energy of the hydrogen-hydrogen distance is compared between that of a stable model (Allegro trained at \qty{600}{K}) and the unstable Allegro model trained at \qty{300}{K}. This comparison indicates that the Allegro \qty{300}{K} model significantly underestimates the repulsive energy between hydrogen atoms, leading to unphysical conformations from which the model fails to recover.

Additionally, we calculate the \caps{PES} of the \NequIP model from\;\cite{forces} using the same procedure of passing a uniform distribution of dihedrals to the model. The results, shown in Figure \ref{fig6}, demonstrate that the \caps{PES} does not have gaps and does not exhibit the flat regions observed in\;\cite{forces}, where calculations were carried out using well-tempered Metadynamics. The observed energy barriers, however, are quite high, reaching $\simeq\qty{200}{kJ.mol^{-1}}$\!. Those flat regions in the \caps{FES} of\;\cite{forces} are now attributed not to the inaccuracy of the model, but rather to the intricacies of the well-tempered Metadynamics method for free energy estimation. As outlined in the original article of well-tempered Metadynamics\;\cite{wtmetad}, the bias factor limits the exploration of energy barriers below the $T+\Delta T$ range. Without a previous estimate of the expected energy barriers in the system, the choice of a small bias factor could be inadequate for overcoming the large energy barriers of the model, leading to limited exploration of the free energy landscape. The energy scale of Fig. \ref{fig6} is approximately 4 times larger than that provided in the classical reference in\;\cite{forces}; we therefore decided to create a new dataset from the one deposited in\;\cite{forcesgithub}, by dividing the forces by 4.184, the factor between kcal and kJ. After this modification, the \caps{PES} of the \NequIP model of\;\cite{forces} is shown in fig. \ref{fig6}b, where the shape and energy scale are closer to the ground truth (fig. \ref{fig6}c). Nevertheless, this modified \NequIP model still exhibits the instability of the original version, and no stable simulations can be achieved using enhanced sampling over the intended 5 ns duration. All tests result in hydrogen atoms detaching from the \caps{ADP} molecule, and despite implementing constraints on hydrogen bonds in another set of free energy calculations, the simulations consistently fail after a few picoseconds.

\subsection{Test 5}
\begin{figure}[htb]
\centering
\includegraphics[width=0.9\textwidth]{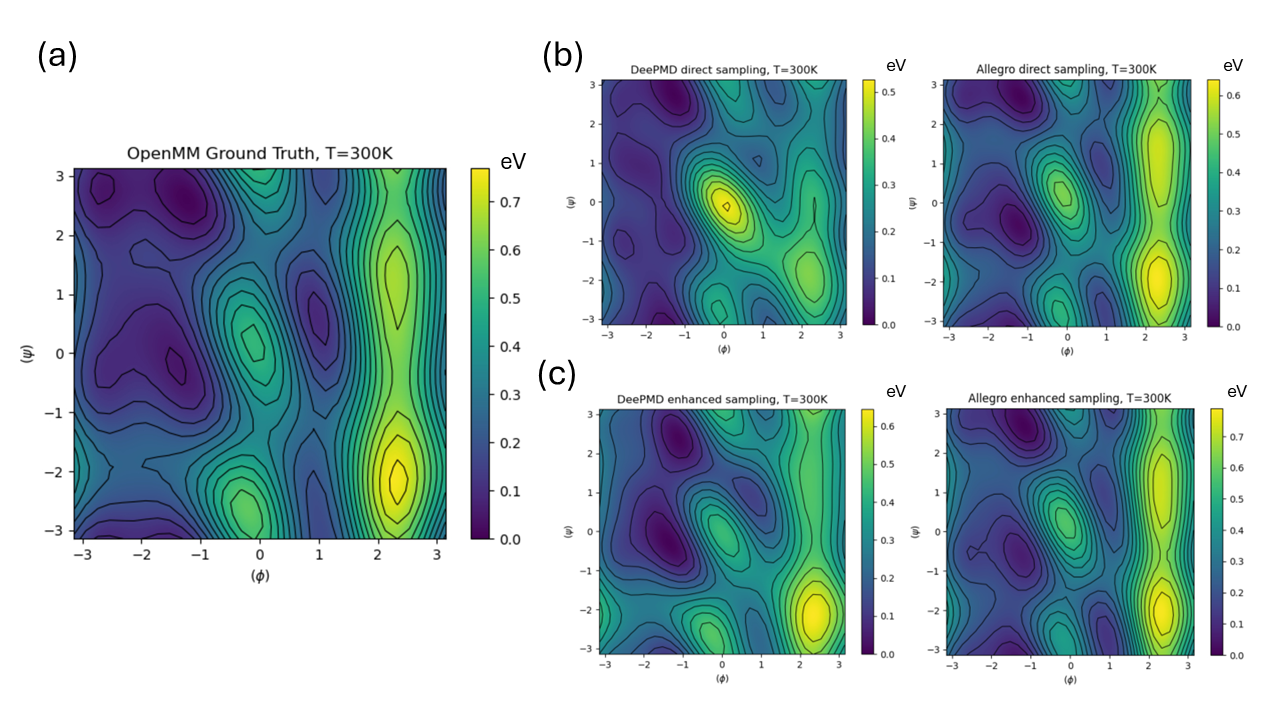}
\caption{Summary of Test 5 results: (a) \caps{FES} of \caps{ADP} in TIP3P water in classical \caps{MD} with full flexible water and \caps{ADP} molecules; (b) the \caps{FES} of \caps{ADP} using the \DeePMD and Allegro models trained on 10,000 snapshots from direct sampling at \qty{600}{K}; (c) \caps{FES} of \caps{ADP} using \DeePMD and Allegro models trained on 10,000 snapshots taken from enhanced sampling at \qty{600}{K}. All \caps{FES} are generated using the Spectral~\caps{ABF} scheme at \qty{300}{K}.}\label{fig8}
\end{figure}

In this last test, we evaluate the capabilities of the \caps{MILP} for handling different local environments for same atom types, since we have two different molecules in the system, water and \caps{ADP}. The results summarized in fig. \ref{fig8} show that both \capsp{MILP} studied in this work, \DeePMD and Allegro, can generate stable potentials with direct and enhanced datasets. The main difference comes in the accuracy of the \caps{FES}, where Allegro exhibits rmse values around five times smaller than those from \DeePMD (see table \ref{table:rmse}).

\section{Discussion}\label{discussion}

The results from Test 1 demonstrate that, contrary to past reports, \capsp{MILP} such as \DeePMD and Allegro can learn a stable \caps{PES} from relatively small datasets derived from unbiased classical simulations for the \caps{ADP} system. We suggest that previous challenges with these frameworks, as reported in\;\cite{forces,consider}, may have been related to the inclusion of mechanical constraints in the training datasets rather than the complexity of the underlying free energy landscape. Additionally, we observe that the internal symmetries within the \caps{MILP} descriptors, particularly the inclusion of equivariance, are crucial for achieving accurate predictions when the model is applied to new, unexplored regions (see Table \ref{table:rmse}). Similar results have been observed in equivariant models of water\;\cite{maxson2024transferable}. For the second test, the use of enhanced sampling techniques solves the accuracy issue observed in Test 1. With enhanced sampling, the \caps{MILP} model gains access to regions that are inaccessible in traditional unbiased simulations. This inclusion of higher energy points in the dataset allows the \caps{MILP} model to reproduce the entire \caps{FES} with very good accuracy. Once again, the equivariance feature proves to be a highly data efficient for learning accurate \capsp{PES} (table \ref{table:rmse}). Moreover, the results using the \caps{DFT} data underscore the essential role of enhanced sampling techniques in generating an accurate \caps{FES}, even when accounting for underlying symmetries in the \caps{MILP} model and the inherent complexity of the free energy landscape. Test 3 illustrates the importance of incorporating both the energy and the forces into the training data set. At this level, implicit solvent models can be effectively learned, with rmse values below \qty{0.01}{eV} (table \ref{table:rmse}), independent from the range of exploration in the \caps{FES}, as long as the potential energy is included into the loss function for the training. For Test 4, which represents the greatest challenge for the \capsp{MILP} explored in this work, as it involves training with only forces, and such forces are a combination of the internal forces of the \caps{ADP} molecule and the forces exerted by the surrounding solvent. In this case, models trained in Allegro using direct or enhanced datasets can maintain stability, but larger datasets are required to achieve good accuracy. This is because each conformation of the \caps{ADP} molecule must be paired with a diverse set of water molecule configurations to capture the full range of solvent forces. The surrounding water molecules play a critical role in this process, as they provide the forces that are essential for the accurate representation of the solvent environment, which is fundamental to the system's overall behavior. The accuracy of the Allegro and \DeePMD models trained with enhanced datasets can be compared that of other approaches, such as those used in\;\cite{implicitw}. In that work, the authors used a multi-window scheme, restraining the conformation of the \caps{ADP} molecule in various regions across the free energy landscape to sample the water force contribution. Our results show that once the mechanical constraints are removed from the datasets, the ability to correctly capture the fast modes of the system dictates the stability in the simulations, rather than the complexity of the slow modes, as was previously inferred\;\cite{forces}. A complex \caps{PES}, such as the dihedral map of \caps{ADP} in solution, can be accurately described by equivariant models, even if they cannot provide stable dynamics. This instability is due to a limited exploration of the fast modes orthogonal to the slow mode \caps{PES} surface as, for example, the distance between the two hydrogen atoms bonded to the nitrogen in the \caps{ADP} molecule (fig. \ref{fig5}). We highlighted the effect of the temperature on the stability of the models trained with data taken from classical \caps{MD} since higher temperatures allow for better exploration of the fast modes of the system.
Finally, for Test 5, we found similar behavior to the results in Tests 3 and 4, i.e, having a combination of energy and forces in the training dataset makes the training more accurate; in this case, both models (\DeePMD and Allegro) can learn a stable and reasonably accurate model of \caps{ADP} and water simultaneously, as long as there are no constraints in both types of molecules in the dataset. Both models can learn the different local environment for oxygen and hydrogen atoms that are part of \caps{ADP} or water molecules, and in combination of using a high temperature (\qty{600}{K}), they can learn the bonded and non-bonded interactions for all the atoms in the system without instabilities. Notably, the equivariant model is shown to be more accurate than the invariant one in both direct and enhanced sampling datasets (table \ref{table:rmse}).

\section{Conclusion}\label{sec12}
In this work, we have highlighted several key features that have interfered with the design of benchmarks for evaluation of \capsp{MILP} using classical \caps{MD} simulations. First, it is important to avoid mechanical constraints that can introduce biases in force calculations, and can significantly influence the configurational space from which \capsp{MILP} learn. Second, we have shown that enhanced sampling leads to improved datasets for training. By employing datasets free from constraints, we demonstrate that the equivariant features embedded in \caps{MILP} models can accurately capture the slow modes of the systems of interest, such as the alanine dipeptide in diverse environments{\emdash}including vacuum, implicit, and explicit solvents{\emdash}tasks that previous studies suggested might be difficult to achieve\;\cite{forces,consider}.
These equivariant symmetry features considerably reduce the minimum size of the necessary training datasets, compared with models that only have invariant descriptors. Additionally, we have shown that the main bottleneck for reaching stability in \caps{MILP} simulations comes from the incomplete exploration of the fast modes of the system, where high temperatures and/or enhanced sampling simulations can effectively alleviate the issue. With these considerations, we hope this work provides a useful framework for future benchmark designs,  guiding efforts to minimize potential biases in the evaluation of \capsp{MILP}.

\section{Methods}\label{sec13}
For the dataset generated for the task 1, \caps{LAMMPS} engine was used. The parameters of the \caps{ADP} model were extracted from the \caps{SSAGES} distribution\;\cite{ssages}. \caps{ADP} is modeled with the Amber \textls[70]{{f}f99SB} force field. The simulation was in the \caps{NVT} ensemble at \qty{500}{K} using the nose-hoover integrator with tau parameter of \qty{100}{fs}. Configurations were collected every \qty{2}{ps} and 1000 snapshots were taken with energy and forces. To generate the ground truth \caps{FES}, \caps{LAMMPS} was coupled to \PySAGES and the integration was done at \qty{300}{K} with the langevin integrator with damping parameter of \qty{1000}{fs}. The Spectral~\caps{ABF} method was used with \qty{5}{ns} of simulation with the default parameters in \PySAGES. The integration grid is 32 by 32 bins. 

For task 2, simulations were done in \caps{ASE} coupled with \caps{SANDER}\;\cite{ambertools}. The langevin thermostat was used with a friction parameter of \qty{10}{ps^{-1}} at \qty{500}{K} of temperature. The enhanced sampling was achieved using Spectral~\caps{ABF} method in \PySAGES with the same parameters as above. Snapshots were collected every \qty{2}{ps} and 1000 frames was saved. For \caps{DFT} single point calculations, \caps{CP2K} was used employing the Gaussian Plane Wave (\caps{GPW}) method. The \caps{GTH-PBE} pseudopotential and the \caps{DZVP-GTH-PBE} basis set were chosen for calculations with the \caps{PBE+D3} functional using a plane-wave cutoff of \qty{550}{Ry}. The direct simulations were done at the same temperature collecting datapoints every \qty{2}{ps}.

For task 3, simulations were done in \caps{ASE} coupled with \caps{SANDER}\;\cite{ambertools} setting \texttt{igb\,=\,1}. The langevin thermostat was used with a friction parameter of \qty{10}{ps^{-1}} at \qty{500}{K} of temperature. The enhanced sampling was achieved using Spectral~\caps{ABF} method in \PySAGES with the same parameters as above. Snapshots were collected every \qty{2}{ps} and 1000 frames was saved. The direct simulations were done at the same temperature collecting datapoints every \qty{2}{ps}.

For task 4, simulations were done with \OpenMM. The langevin thermostat was used with a friction parameter of \qty{10}{ps^{-1}}\!. The enhanced sampling was achieved using Spectral~\caps{ABF} method in \PySAGES at \qty{300}{K} of temperature. To remove any enhanced sampling forces that comes from \PySAGES, the trajectories are passed to \OpenMM again to serve as single point calculation, removing the bias forces incorporated from \PySAGES. For direct simulations, \qty{600}{K} was fixed for the temperature.  Datasets with 1000 snapshots were sampled every \qty{2}{ps} for direct simulations and \qty{10}{ps} for enhanced sampling. For 40,000 snapshots datasets, the recollection of configurations was every \qty{2}{ps} in enhanced and direct simulations.

For task 5, simulations were done with \OpenMM. The langevin thermostat was used with a friction parameter of \qty{10}{ps^{-1}}\!. The \caps{SETTLE} constraints in \caps{TIP3P} water model were removed from the simulations and the Amber \textls[70]{{f}f99SB} force field was used for \caps{ADP}. The enhanced sampling was achieved using Spectral~\caps{ABF} method in \PySAGES at \qty{600}{K} of temperature. To remove any enhanced sampling forces that comes from \PySAGES, the trajectories are passed to \OpenMM again to serve as single point calculation, removing the bias forces incorporated from \PySAGES. For direct simulations, \qty{600}{K} was fixed for the temperature.  Datasets with 10000 snapshots were sampled every \qty{2}{ps} for both datasets.

Detailed scripts for running and training all tests are available in\;\cite{depablogithub}. 

\backmatter


\bmhead{Acknowledgements}
This work is supported by the Department of Energy, Basic Energy Sciences, Materials Science and Engineering Division, through the Midwest Integrated Center for Computational Materials (\CAPS{MICC}{o}\caps{M}), and by the National Science Foundation under Grant No. 2022023. The authors also acknowledge the Research Computing Center of the University of Chicago for computational resources.


\noindent

\bigskip
\bibliography{bibliography}
\begin{appendices}

\section{Supplementary figures}\label{secA1}
\counterwithin{figure}{section}

\begin{figure}[htb]
\centering
\includegraphics[width=0.9\textwidth]{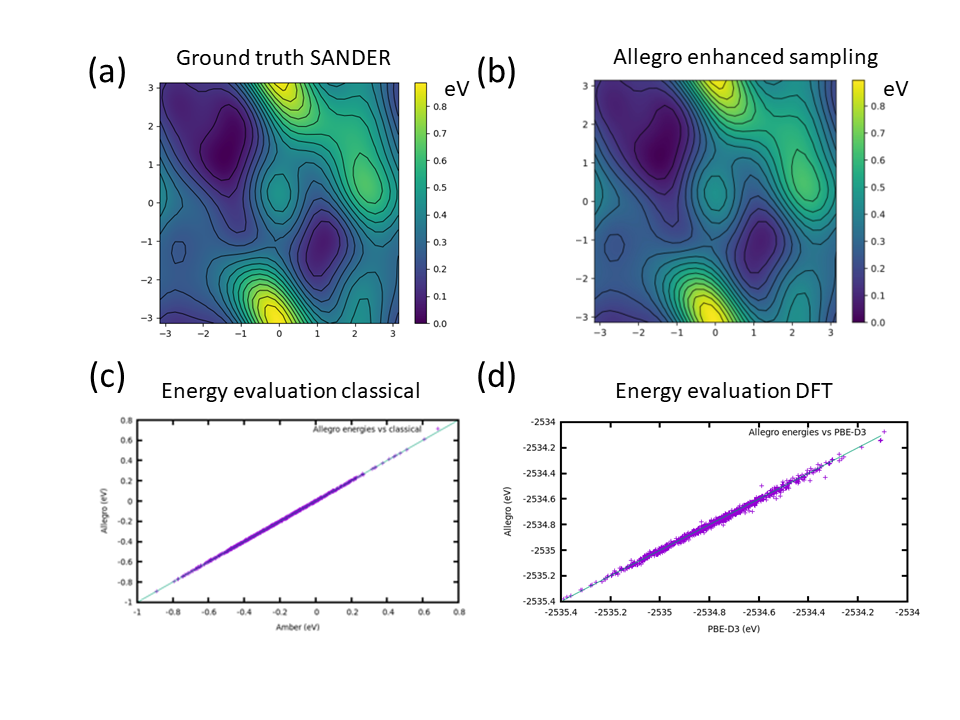}
\caption{Uniform training of \caps{ADP} at classical and \caps{DFT} level. a)Free energy surface of classical \caps{MD} in \caps{SANDER}. b) Free energy of \caps{ADP} with Allegro with uniform sampling. c) Energy evaluation of the classical vs Allegro model. d) Energy evaluation of the \caps{DFT} vs Allegro model. All free energies maps were generated using the Spectral~\caps{ABF} method in \PySAGES. Energy evaluation was done with 1000 snapshots taken from the production runs of \PySAGES.}\label{fig2}
\end{figure}

\begin{figure}[htb]
\centering
\includegraphics[width=0.9\textwidth]{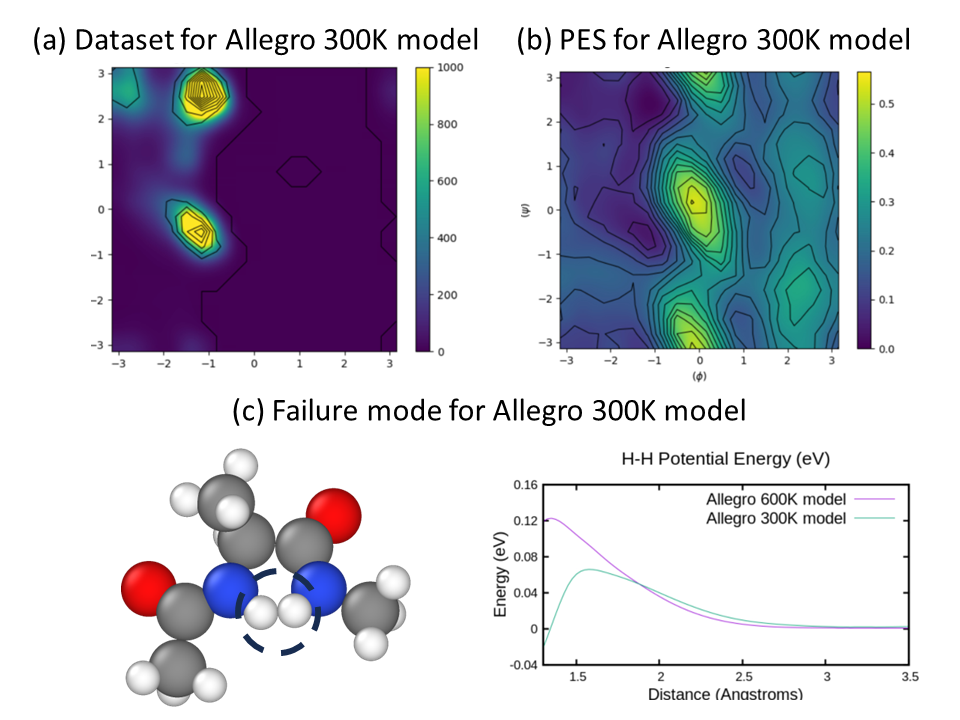}
\caption{Importance of temperature in training dataset for model stability. a) Dataset distribution for \caps{ADP} conformations in explicit water generated by unbiased sampling at \qty{300}{K}. b) Allegro \qty{300}{K} model \caps{PES} surface generated by averaging the potential energy over a 20 x 20 grid covering the dihedral landscape. c) Analysis of the failure mode of the Allegro \qty{300}{K} model. In simulations, the hydrogen-hydrogen distance goes too close, so nonphysical conformations are generated and this lead to molecule collapse. Analysis of the potential energy of the interaction between hydrogens in Allegro \qty{300}{K} and allegro trained with \qty{600}{K}.}\label{fig5}
\end{figure}

\begin{figure}[htb]
\centering
\includegraphics[width=0.9\textwidth]{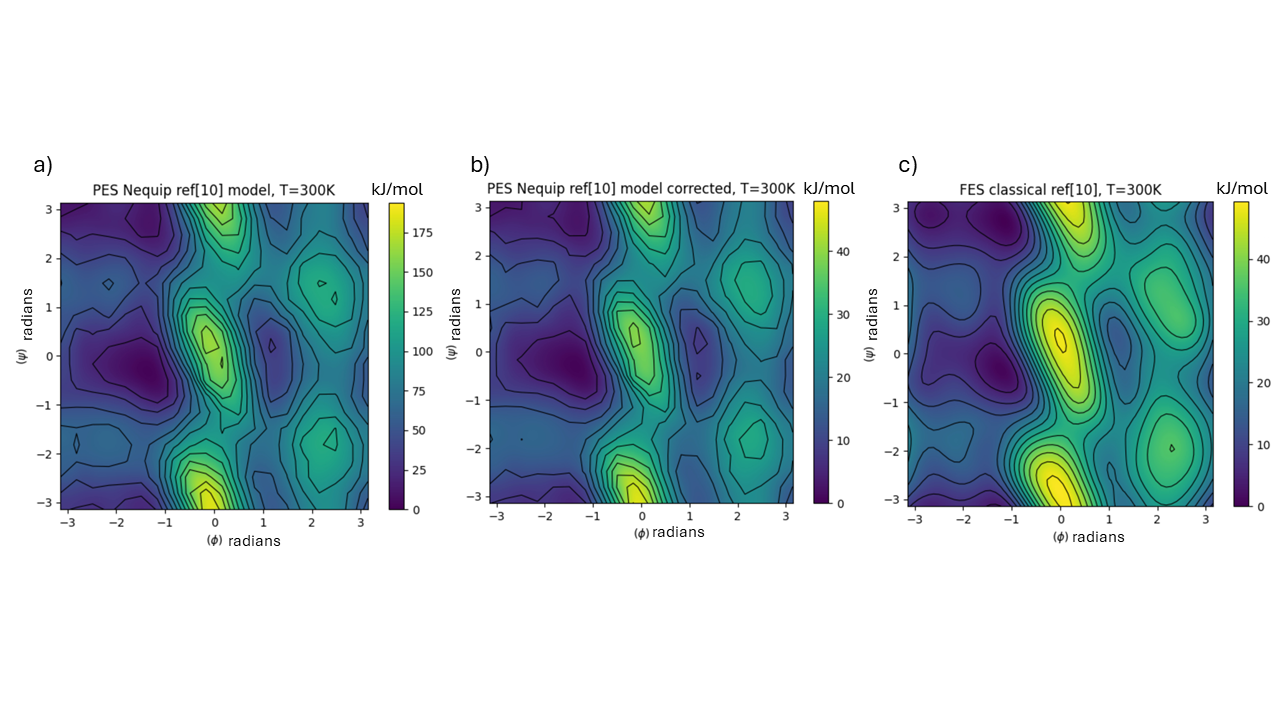}
\caption{Importance of uniform sampling for uncover model accuracy and stability. a) \caps{PES} of original \NequIP model training on data from\;\cite{forces}. Note that the energy is around 4 times larger than reference c). b) \caps{PES} of \NequIP model of\;\cite{forces} once the proper energy conversion factor is used. c) \caps{FES} of the reference system in\;\cite{forces}. \caps{PES} were obtained by passing an 80,000 snapshots uniformly distributed on the dihedrals 2D map and the potential energy is calculated on a 20 by 20 grid. The \caps{FES}, the training parameters, and the training data are obtained from the supplementary information of\;\cite{forces,forcesgithub}.}\label{fig6}
\end{figure}

\begin{figure}[htb]
\centering
\includegraphics[width=0.9\textwidth]{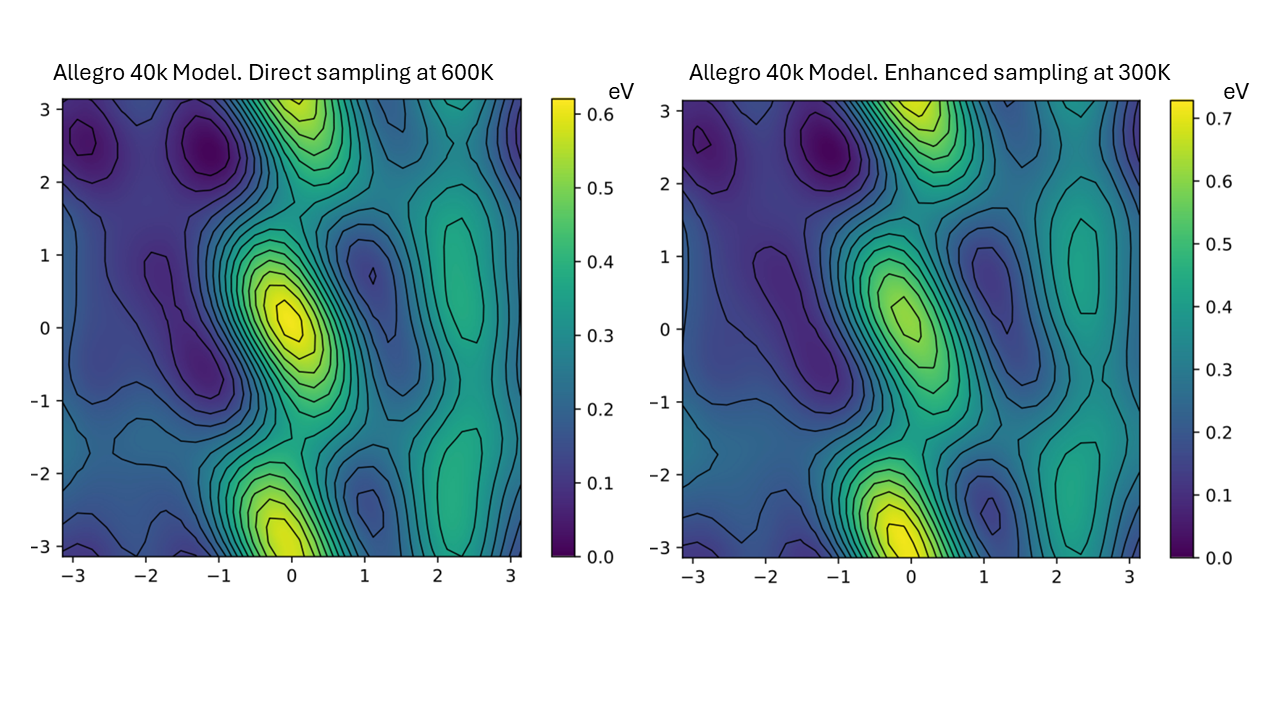}
\caption{Increasing dataset for Test 4. \caps{FES} calculated with 40000 snapshots with direct sampling at \qty{600}{K} and with enhanced sampling at \qty{300}{K}. The \caps{FES} was calculated using Spectral~\caps{ABF}.}\label{fig7}
\end{figure}

\begin{table}[htb]
\caption{Root Mean Squared Error on \caps{FES} results for type of model on each test. The first 3 tests are reported for model trained with 1000 frames. For \caps{DFT} results, the reported \caps{RMSE} is respect to potential energy from the 1000 snapshots taken from the enhanced sampling simulation. Test 4 results are reported with 40,000 snapshots of training. Test 5 results are reported from 10,000 snapshots of training. The reporting units are in eV. If the model was not able to run the dynamics, it is reported as {---}. The table shows that Allegro, an equivariant neural network, consistently outperforms \DeePMD. It is evident from the \caps{RMSE} values, where Allegro achieves an order of magnitude lower \caps{RMSE} than \DeePMD, which also encounters failed cases in running the dynamics.}

\begin{center}
\begin{small}

\sisetup{table-format = 1.5}
\newcolumntype{D}{>{\lsstyle\osf}S}

\begin{tabular}{>{\scshape}lDD}
\toprule
{\upshape Test} & {Allegro (eV)} & {\DeePMD (eV)} \\
\midrule
1. Classical Direct & 0.0920 & 0.2241 \\
1. \caps{DFT} Direct  & 0.23225 & {---} \\
2. Classical Enhanced & 0.00542 & {---} \\
2. \caps{DFT} Enhanced & 0.0155 & {---} \\
3. Implicit Direct & 0.00988 & {---} \\
3. Implicit Enhanced & 0.00769 & {---} \\
4. Explicit Direct 40k & 0.04451 & 0.11314 \\
4. Explicit Enhanced 40k & 0.01153 & 0.11438 \\
5. Full Direct 10k & 0.06627 & 0.27799 \\
5. Full Enhanced 10k & 0.06060 & 0.32002 \\
\bottomrule
\end{tabular}
\vspace{-0.2in}
\label{table:rmse}

\end{small}
\end{center}
\end{table}



\end{appendices}


\end{document}